# Highly Nonlinear Ising Model and Social Segregation

M.A. Sumour[1], M.A. Radwan[1], M.M. Shabat[2],
[1] Physics Department, Al-Aqsa University, P.O.4051, Gaza, Gaza Strip,
Palestinian Authority, e-mail msumoor@yahoo.com , ma.radwan@alaqsa.edu.ps.
[2]Physics Department, Islamic University, P.O.108, Gaza, Gaza Strip, Palestinian Authority, e-mail : shabat@iugaza.edu.ps

**Keywords:**
Ising Model, Modified Ising model, statistical physics, urban segregation.

## Abstract:
The usual interaction energy of the random field Ising model in statistical physics is modified by complementing the random field by added to the energy of the usual Ising model a nonlinear term $S^n$. where S is the sum of the neighbor spins, and n=0,1,3,5,7,9,11.
Within the Schelling model of urban segregation, this modification corresponds to housing prices depending on the immediate neighborhood. Simulations at different temperatures, lattice size, magnetic field, number of neighbors and different time intervals showed that results for all n are similar, expect for n=3 in violation of the universality principle and the law of corresponding states. In order to find the critical temperatures, for large n we no longer start with all spins parallel but instead with a random configuration, in order to facilitate spin flips. However, in all cases we have a Curie temperature with phase separation or long-range segregation only below this Curie temperature, and it is approximated by a simple formula: Tc is proportional to 1+m for n=1, while Tc is roughly proportional to m for n >> 1 .

## Introduction:

The Ising model is the simplest and most famous spin system model to study phase transitions. It was introduced in 1925 by Ernst Ising in his Ph.D. thesis. He solved the model completely for one dimension, and found that no phase transition occurs. He concluded that this should be the case for all dimensions, a fatal mistake. Inspite of this, the model was renamed after him.

In the Schelling – Ising model of urban segregation [1-2] all lattice sites are equivalent. In reality, some houses are cheap and others are expensive. And usually of two groups in a population, one is poorer than the other.

We study the urban segregation by these models which use two groups A and B of people distributed on a square lattice, with group A corresponds to up spins ↑ (+1) and group B to down spins ↓ (-1), in a random magnetic field on finite samples, to check cheap and expensive residences.

Urban segregation in this article will be studied by using Statistical Physics, when the usual interaction energy of the random field Ising model is modified by adding to the random field an odd power of the sum of the neighbor spins. Within the Schelling model of urban segregation, this modification corresponds to housing prices depending on the immediate neighborhood.

The general equation of the random-field Ising model is given by :

$$E = -J \sum s_i s_k - \sum h_i s_i \qquad (1)$$

where $s_i$ is the spin (±1), J the interaction constant, $h_i$ a random field, and $S_i = \sum_k s_k$ with the sum over all nearest neighbors of i.

This random field Ising energy can be rewritten as:

$$E = -\sum_i s_i \left(JS_i + h_i\right) \quad (2)$$

In the Ising model without the random field in two and more dimensions, two neighboring spins have due to their interaction $-J\, S_i\, S_k$ a higher probability to belong to the same group than to belong to two different groups. If the difference between these two probabilities is large enough, $T < T_c$, domain sizes can grow to infinity in an infinite lattice, while only small clusters are formed for smaller differences in the probabilities, $T > T_c$ [2] . These probabilities - controlled through $-J/\, k_BT$- lead to these different regimes, separated by a sharp phase transition at $T = T_c$, not obvious from the definition of the interaction J S_i S_k , physicists took many years to find it, and it is typical of complex systems.

The earlier standard Ising model gives results similar to the properly modified Schelling model [3-8].

The magnetic field is considered to refer to the price of residence: Group A prefers to go to the cheap residences and group B to the expensive residences. This can be simulated by a random magnetic field which is + h on half of the places (attracting people of group A = up spins) and is −h on the other half of the places (attracting group B = down spins). The signs of the field are distributed randomly, and the model is called the random-field Ising model.

In the spin 1/2 Ising model with Glauber kinetics on the square lattice, we interpret the two spin orientations as representing two groups of people. Nearest neighbors are coupled ferromagnetically, i.e. people prefer to be surrounded by others of the same group and not of the other group. Starting with random initial distribution of zero magnetization (= number of one group minus number of the other group), we check if "infinitely" large domain are formed. It is well known that they do so for $0 < T < T_c$ where $T_c = 2.269$ is the critical temperature in units of the interaction energy for the usual Ising model.

The Glauber kinetics is simulated on the computer by flipping a spin if and only if a random number between 0 and 1 is smaller than the probability which is equal $\exp(-\Delta E/k_BT)/[1 + \exp(-\Delta E/k_BT)]$, where $\Delta E$ is the energy change produced by this spin flip.

## Ising model in Sociophysics:

If two groups of people A and B, rich and poor, are distributed on a square lattice, we use the Ising model with random magnetic field at small and large temperatures for times up to 9000. The social meaning of temperature T is not what we hear in weather reports but an overall approximation for all the more or less random events which influence our decisions.

Thus the temperature here can have two meanings: tolerance and noise. Tolerance means that for high T one is willing to live among neighbors from a different group, a and low T means that people strongly prefer to live among neighbors of their own group. The alternative interpretation is T = noise; T then measures all those facts of life outside the model which force people to move to another residence even though they like their old residence better.

Within the Ising model of urban segregation, our modification corresponds to housing prices depending on the immediate neighborhood. Now we add to the J-term in equation (2) the energy J'(S_i)$^n$, where J' is another interaction constant and n = 1,3,5,7,9,11,… some odd integer.

For n=1 we recover the usual Ising model with J+J' instead of only J as interaction constant. Thus with n=1 we can test whether our program gives correct results, and with n = 3,5,7,9,11 we can get results which to our knowledge are new. By n=0 we denote the standard Ising model with J only, setting J' to zero, while otherwise at first J'=J, later we use also m= J'/J up to 5. With even n (n=2,4,6...) the new term $S_i^n$ is always positive which seems less interesting.

Thus the energy of the new modified Ising model becomes

$$E = -\sum_i s_i \left( J\, S_i + J'\, S_i^{\,n} + h_i \right), \quad S_i = \sum_k S_k \qquad (3)$$

k = neighbor of site i . Similar models were studied earlier with [10-11]
In physics we can imagine that the strength of the magnetic dipole moment of an atom is influenced in a nonlinear way by its neighbour atoms.

## Discussion and results:

In the present work we use the simple standard two-dimensional Ising model with Glauber dynamics instead of the complicated Schelling model. We initially carry out our simulation at temperature of T = 2.0, with size of square lattice of 500×500, at time = 4000, and a field value = 0.1.We noticed large domains at this temperature and this field as in figure 1, but when the field is changed from 0.1 to 0.9 with the same parameters the domains decrease and small domains can be noticed, as in figure 2. Simulations at high temperature T = 17 with the same parameters showed that the two populations mix without any large clusters as in figure 3.

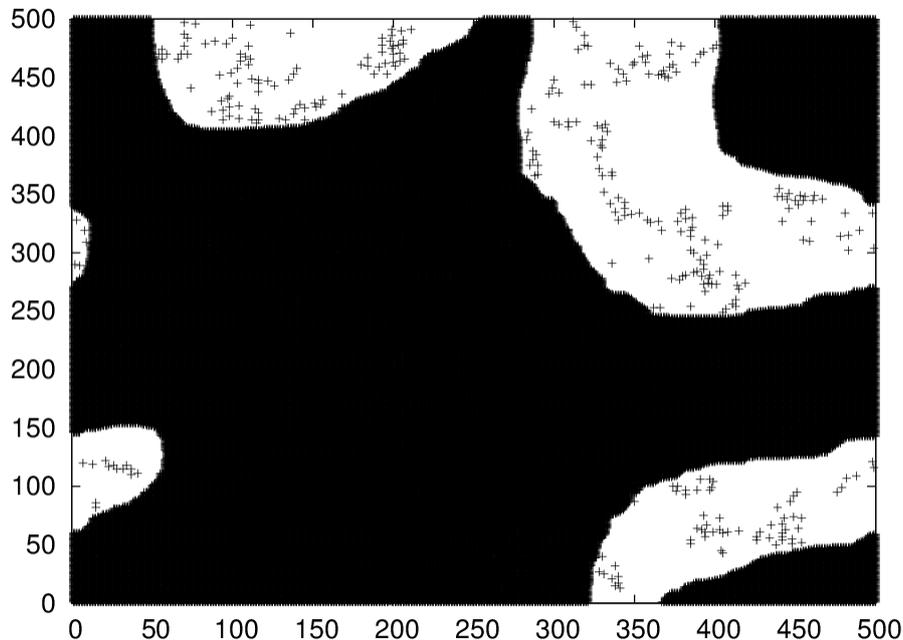

Figure(1): Shows the large domains of groups in small random field of + 0.1, and low temperature of 2.0, at time equal 4000, L=500 and n=3.

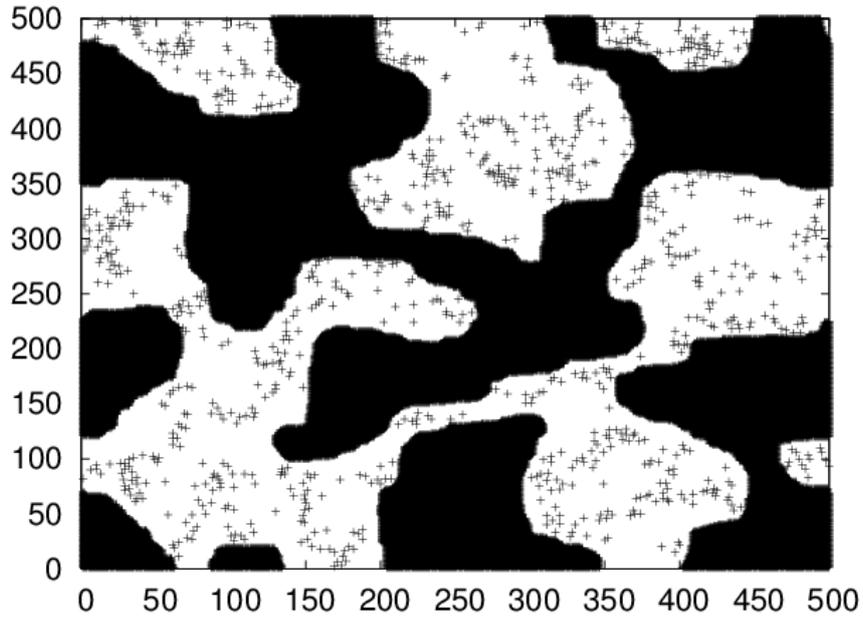

Figure(2): Shows the decrease of domains of groups as h=+ 0.9, and low temperature of 2.0, at time equal 4000, L=500 and n=3.

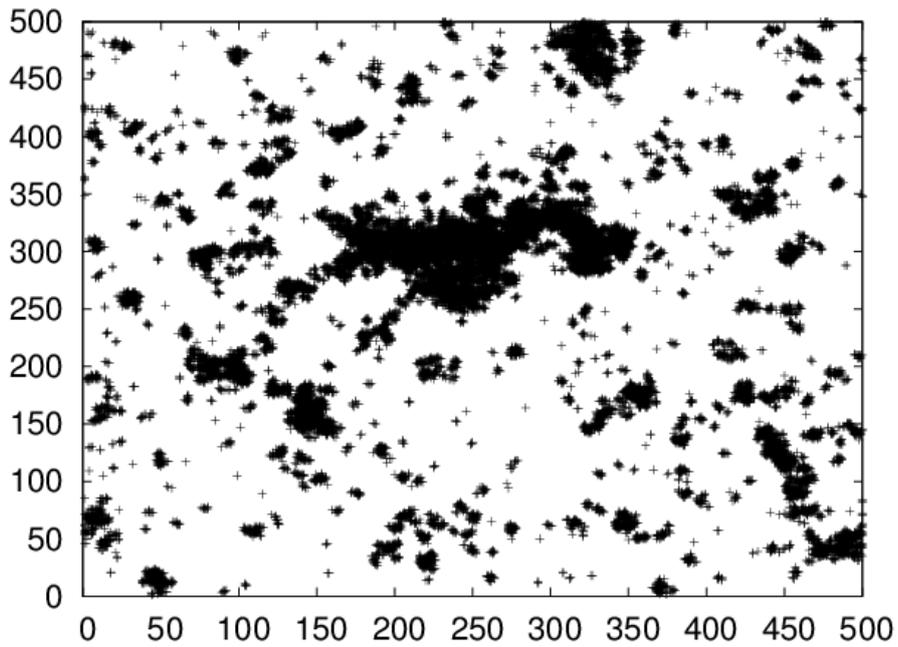

Figure(3): Shows the two populations mix without any large clusters. At h= + 0.9, and high temperature of 17, at time equal 4000, L=500 and n=3.

It was also demonstrated that there are no growing domains obtained at high field, while it was observed that the magnetization of the model changes from ferromagnetic to paramagnetic at the random field values higher than 0.256 for n=1,3,5,7,9,11.

In our simulation at different temperatures we check the critical temperatures Tc for different n=0,1,3,5,7,9,11. First we set for n=0, size of lattice 2000, time =9000, and put the value of the field=0.001. We find the critical temperature = 2.26 which agrees with the known critical temperature for the normal Ising model n=0.

Figure 4 shows the magnetization versus time where T=Tc in the middle line, the lower line belongs to T > Tc , and the higher line to T > Tc.

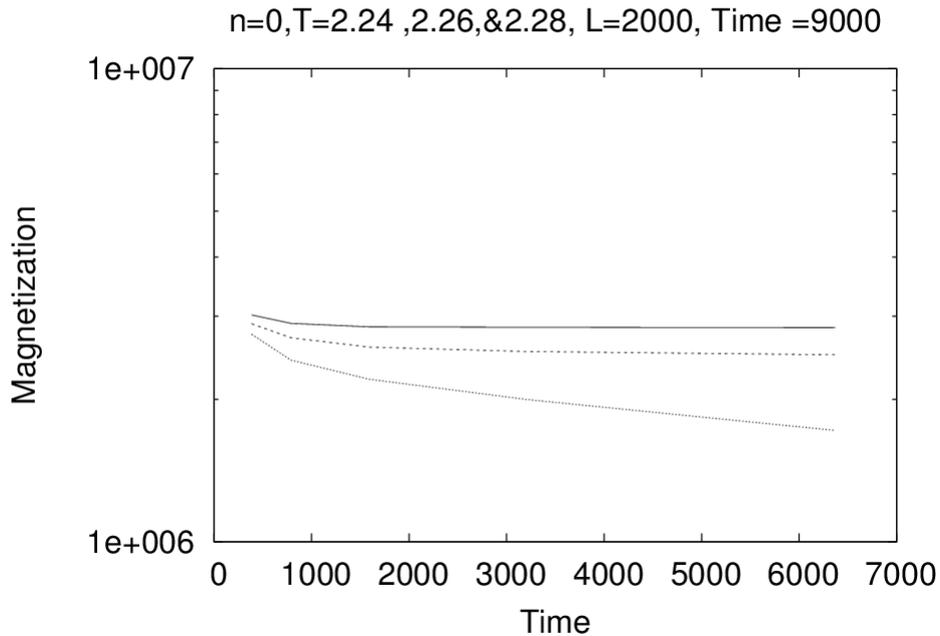

**Figure (4)** : Magnetization versus time with temperature = 2.24(up), 2.26(middle) critical temperature, and 2.28 (down).

Similarly we go to n=1,3,5,7,9,11 for the same size of lattice L=2000,field=0,and time=9000, with different m and we take initially 1% spins down and 99% spins up at different temperatures, in this way the system can more easily move away from the initial configuration and find it's equilibrium, changing the initial fraction of down spins from 0.01 to 0.001 and 0.1 makes no difference in the calculated magnetization, critical temperatures for all n with different m= J'/J listed in table 1:

**Table(1) :Tc for different n and m, at h=0**

| n | 1 | 3 | 5 | 7 | 9 | 11 |
|---|---|---|---|---|---|---|
| m | Tc | | | | | |
| 1 | 4.6 | 17.7 | 38.5 | 148 | 587 | 2350 |
| 2 | 6.9 | 32.5 | 76 | 290 | 1180 | 4710 |
| 3 | 9.2 | 47 | 112 | 442 | 1770 | 7090 |
| 4 | 11.5 | 61 | 149 | 590 | 2360 | 9415 |
| 5 | 13.8 | 76 | 186 | 737 | 2950 | 11750 |

Then we plot figures for each n to see the behavior of the magnetization versus time for n=1,3,5,7,9,11, we show for T=Tc the middle line, the lower line belongs T>Tc, and the higher line to T < Tc as figures(5-8) listed below for m=1 and selected n:

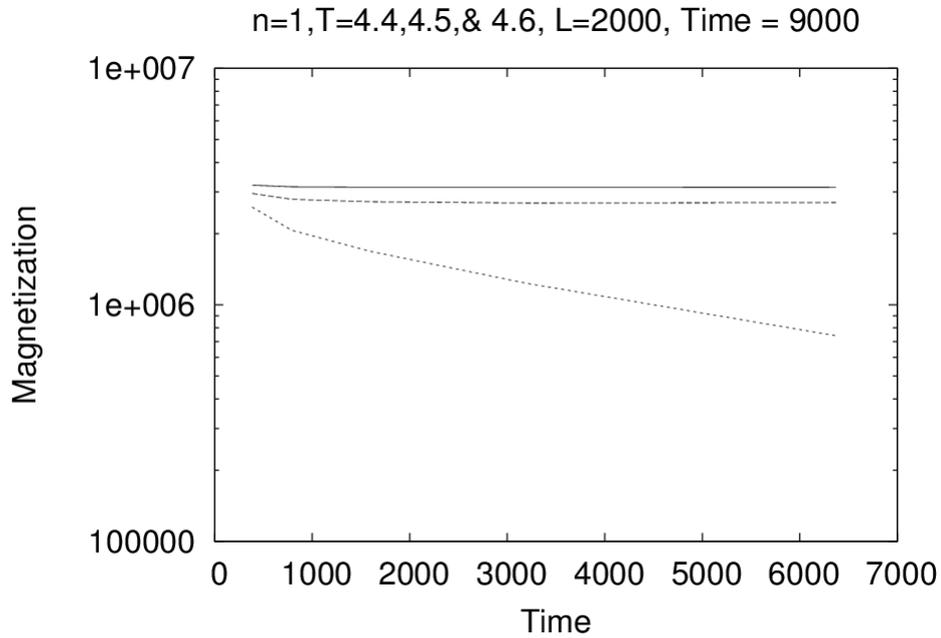

**Figure (5)** : Magnetization versus time with temperature = 4.4(up), 4.5(middle) critical temperature, and 4.6 (down)

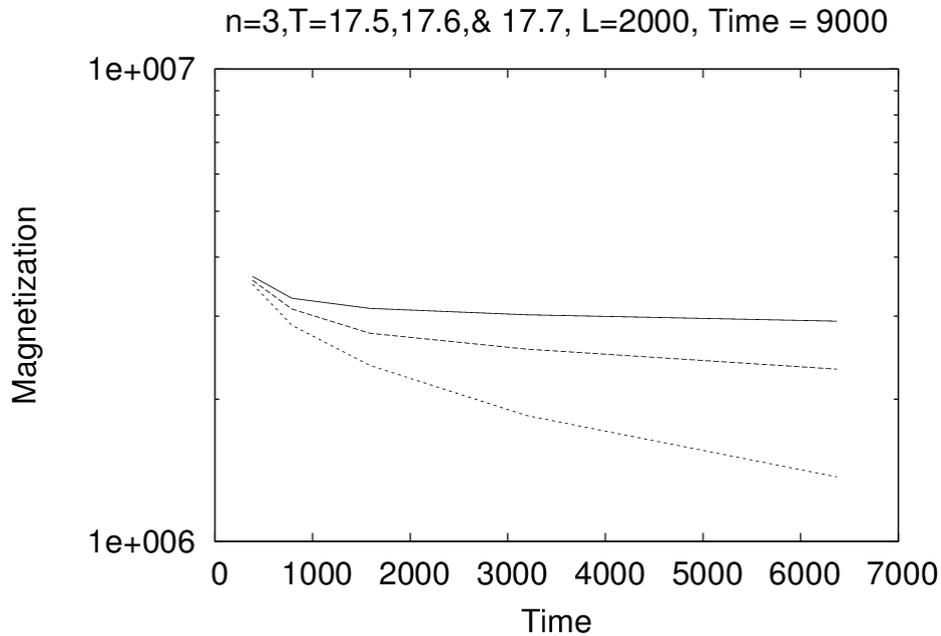

**Figure (6)** : Magnetization versus time with temperature = 17.5(up), 17.6(middle) critical temperature, and 17.7( below ).

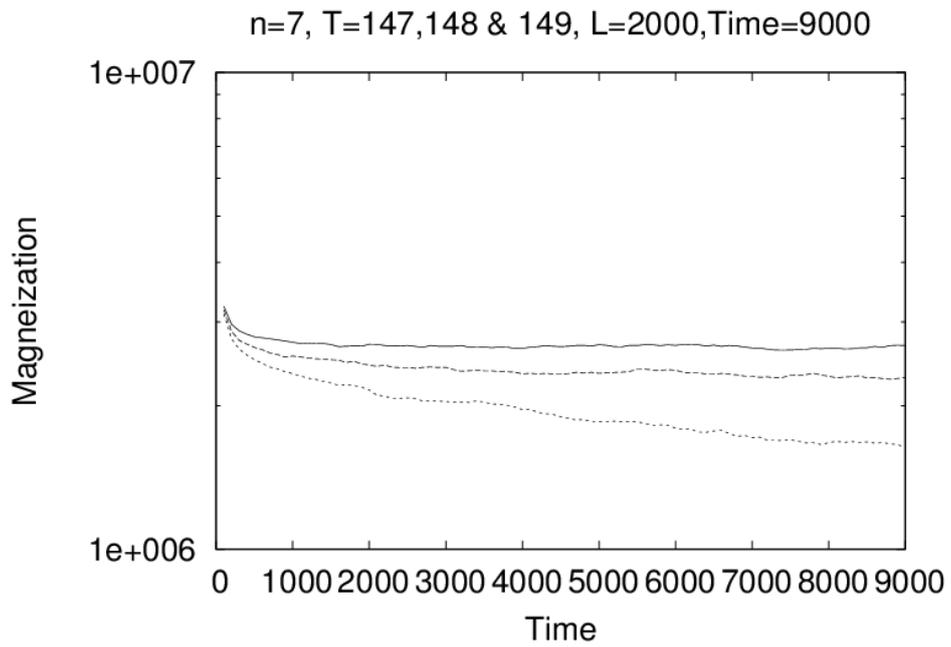

**Figure (7)** : Magnetization versus time with temperature = 147(up), 148(middle) critical temperature, and 149( below ).

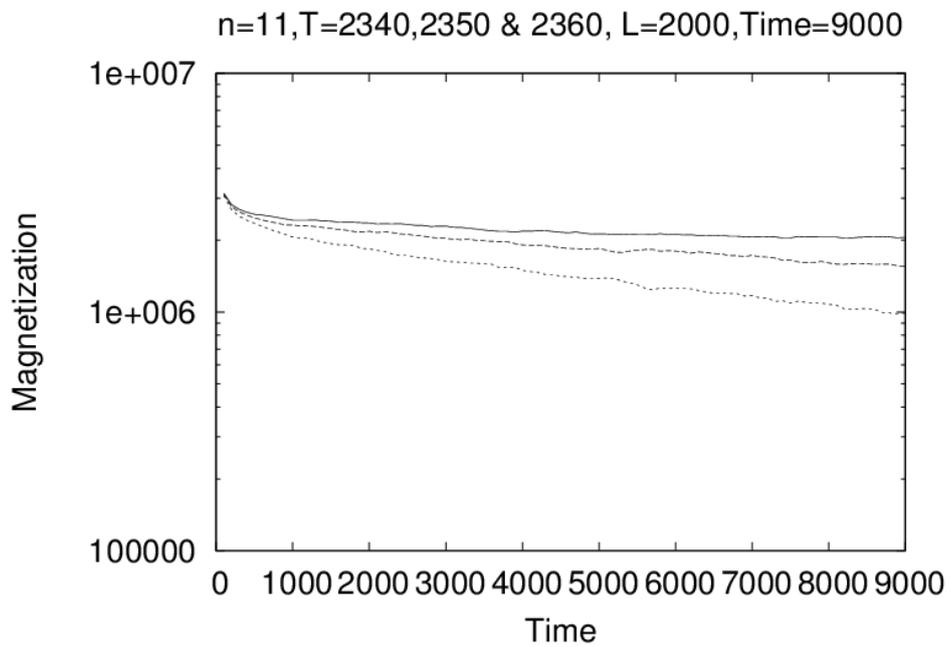

**Figure (8)** : Magnetization versus time with temperature = 2340(up), 2350(middle) critical temperature, and 2360( below ).

Finally we had shown numerically that for n=1 Tc is exactly proportional to 1+m, which is in accordance with the energy equation ( eqn. 3) since the energy is proportional to (S+m*S), while for n >> 1 Tc is roughly proportional to m, since the term S^n dominates the energy, so the term S can be neglected leading to Tc roughly proportional to m.

And at a fixed m, for large n Tc varies exponentially with n.

We get the magnetization, normalized by the number of sites, from our simulation and plot it versus T/Tc, at 0.5 Tc to Tc for each n , size of lattice 2000, time =9000, and put the value of the field=0.001; thus we get figure 7.

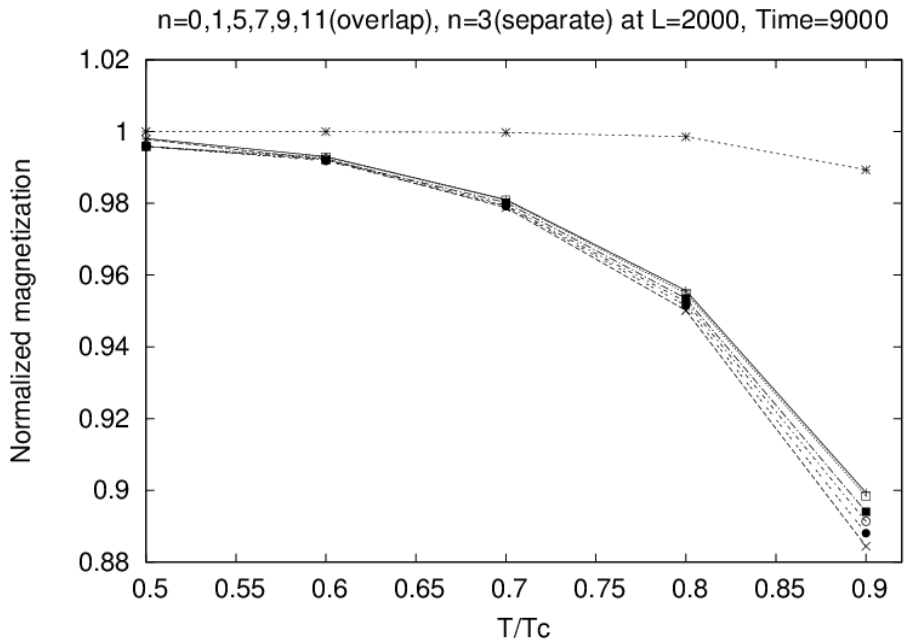

**Figure (9)** : Normalized magnetization versus T/Tc for 0.5Tc to Tc for each n.

But we see that for n= 3 only does not agree with all values of n=0,1,5,7,9,11, and it's deviation from the others we don't understand.

Finally we plot Tc at different strengths J'/J=m for each n as in figure10.

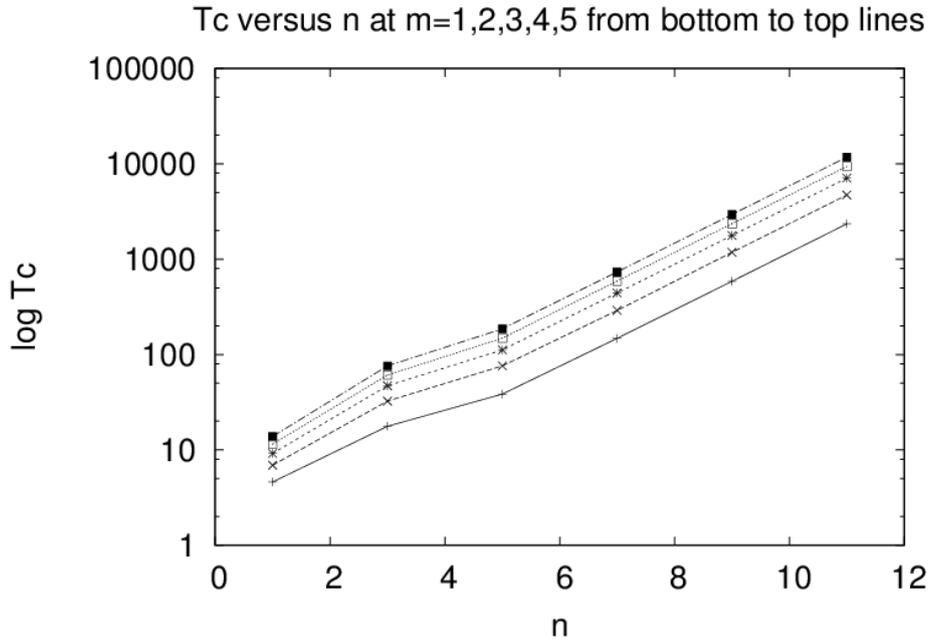

**Figure (10)** : Critical temperatures versus n for m=1,2,3,4,5.

From this figure we find that for a fixed m, for large n Tc varies exponentially with n, i.e.
log Tc = a n + constant, where a is the slope, which agrees with an Arrhenius law.

## Conclusions:

■ We assumed that residences are either cheap or expensive, randomly distributed over the square lattice, and that two groups of people, rich and poor, make up the population. We found that for small fields after a long time the domains are larger than for large fields, in this random-field Ising model of urban segregation. Housing price differences do not prevent segregation if they are not very large.

■We conclude that n=0 and n=1,5,7,9,11 are similar while n=3 differs, in violation of the universality principle and the law of corresponding states. However, in all cases we have a Curie temperature with phase separation or long-range segregation only below this Curie temperature.

■We found numerically that for n=1 Tc is exactly proportional to 1+m, while for n >> 1 Tc is roughly proportional to m. And at a fixed m, for large n Tc varies exponentially with n.

## Acknowledgment

The authors thank Prof. D. Stauffer for many valuable suggestions and fruitful discussions during the development of this work.